\documentclass{article}
\usepackage{spconf, graphicx, booktabs, kotex}
 
\title{Exploring Lexicon-Free Modeling Units for End-to-End Korean and \\
Korean-English Code-Switching Speech Recognition}
\name{Jisung Wang, Jihwan Kim, Sangki Kim, Yeha Lee}
\address{VUNO Inc.\\ 507, Gangnam-daero, Seocho-gu, Seoul}

\begin{document}
\maketitle

\begin{abstract}
As the character-based end-to-end automatic speech recognition (ASR) models evolve, the choice of acoustic modeling units becomes important. 
Since Korean is a fairly phonetic language and has a unique writing system with its own Korean alphabet, it's worth investigating modeling units for an end-to-end Korean ASR task. 
In this work, we introduce lexicon-free modeling units in Korean, and explore them using a hybrid CTC/Attention-based encoder-decoder model.
Five lexicon-free units are investigated: Syllable-based Korean character (with English character for a code-switching task), Korean Jamo character (with English character), sub-word on syllable-based character (with sub-word in English), sub-word on Jamo character (with sub-words in English), and finally byte unit, which is a universal one across language.
Experiments on Zeroth-Korean (51.6 hrs) and Medical Record (2530 hrs) are done for Korean and Korean-English code-switching ASR tasks, respectively.
Sequence-to-sequence learning with sub-words based on Korean syllables (and sub-words in English) performs the best for both tasks without lexicon and an extra language model integration.
\end{abstract}

\section{Introduction}
\label{sec:intro}
Sequence-to-sequence learning with attention-based models has becoming increasingly popular for automatic speech recognition (ASR) \cite{state_of_the_art_seq2seq}, \cite{comparison_seq2seq}, \cite{las}.
Such end-to-end methods directly predict character-based units, which allow us to build ASR systems easily without a hand-designed lexicon.
Accordingly, various acoustic units in sequence-to-sequence models have been investigated on English ASR task, including lexicon-free units such as graphemes \cite{state_of_the_art_seq2seq}, word-pieces \cite{jp_ko_voice_search}, and sentence-pieces \cite{comparative_study_on_transformer_vs_rnn} as well as lexicon-related units such as context dependent (CD) states and context independent (CI) phonemes \cite{analysis_of_attention}.
These studies showed that the choice of modeling units is important in a sequence-to-sequence model.

Development of the character-based end-to-end models can give an advantage especially on Korean ASR tasks, since there is no standard Korean phoneme set and lexicon like CMU dictionary in English.
In the Korean writing system, there are basic letters called Jamo, each representing a consonant or a vowel just as there are alphabet letters in English.
The Jamo letters are combined forming into a syllable block, which is a basic Korean character.
Thus, lexicon-free modeling unit for Korean ASR can be either a Jamo or a syllable-based Korean character.
Furthermore, we applied SentencePiece \cite{sentence_piece} algorithms to generate sub-word units.
In summary, the investigated units in this work are Jamo, syllable, Jamo-based sub-word, syllable-based sub-word.
And we also compared them with byte unit introduced in \cite{bytes_are_all_you_need}, leading to total five different modeling units.
We used Zeroth-Korean benchmark sets, which consists of only Korean words.
As English words as much as Korean words are spoken in a real-world situation, we further performed experiments on a large Korean-English code-switching corpus called Medrec in order to observe units' scalability.
Note that there is no need to build a shared phoneme set, lexicon or language model for a bilingual code-switching ASR system.
The hybrid connectionist temporal classification (CTC)/attention based-decoder model was chosen as the main architecture.
We performed multi-objective training and joint decoding.
Among five units, syllable-based sub-word model showed the best performance in the Korean ASR task.
For a Korean-English code-switching task, the combination of syllable-based sub-word unit and English sub-word unit has achieved the highest performance for Korean-English code-switching tasks.

\section{Relation to prior work}
\label{sec:prior}
Studies on modeling units in sequence-to-sequence learning \cite{comparison_seq2seq}, \cite{analysis_of_attention}, \cite{bytes_are_all_you_need},  \cite{on_the_choice_of_modeling_unit}, has been extended to other language such as Mandarin Chinese: CI-phonemes, syllables (pinyins with tones), Chinese characters, words and sub-words \cite{madarin_comparison_of_modeling_units}, \cite{mandarin_syllable_based_seq2seq}, \cite{mandarin_comparable_study_of_modeling_units}.
However, few related studies has been done on the Korean ASR task.
Accordingly, our work has the following novelties. 
We firstly investigate the lexicon-free acoustic modeling units suitable for an end-to-end Korean ASR task.
We further validate the results' scalability by training on a larger Korean-English code-switching data.

\section{System overview} \label{sec:sys}
\subsection{Model architecture} \label{ssec:sys_architecture}
The architecture used in this work is based on Listen, Attend, and Spell (LAS) model \cite{las}.
LAS consists of a Recurrent Neural Network (RNN)-based encoder, an attention module, and an RNN-based decoder.
The encoder processes the input acoustic feature sequence into high level representations.
The attention module calculates a single representative feature vector for each decoding step.
And the decoder outputs a probability distribution over a character sequence conditioned on the previously predicted labels.
In this work we follow the study \cite{advances_in_joint_ctc_attention} and incorporate a Convolutional Neural Network (CNN) in the encoder network, which includes 4 convolution and 2 maxpooling layers as follows:
2D-Conv (\(ch_{in}=1\), \(ch_{out}=64\), filter sz\(=3 \times 3\)), 2D-Conv (\(ch_{in}=64\), \(ch_{out}=64\), filter sz\(=3 \times 3\)), Maxpooling (kernel \(=2 \times 2\), stride \(=2 \times 2\)), 2D-Conv (\(ch_{in}=64\), \(ch_{out}=128\), filter sz\(=3 \times 3\)), 2D-Conv (\(ch_{in}=128\), \(ch_{out}=128\), filter sz\(=3 \times 3\)), Maxpooling (kernel \(=2 \times 2\), stride \(=2 \times 2\)).\\
Note that the two maxpooling layers in CNN downsamples the input to ratio \((1/2)^2 = 1/4\) along both time and frequency axises.
For an RNN part in the encoder, we use 5-layer Bi-directional LSTM (BLSTM) \cite{lstm} with 512 cells and each BLSTM layer is followed by linear projection layer.
For attention module, location-aware mechanism is used with 512 dimension, 10 for a convolution channel size, and 100 for a filter size.
The decoder network also consists of 2-layer LSTM with 512 cells.
\begin{figure}[ht]
    \centering
        \includegraphics[width=6.7cm]{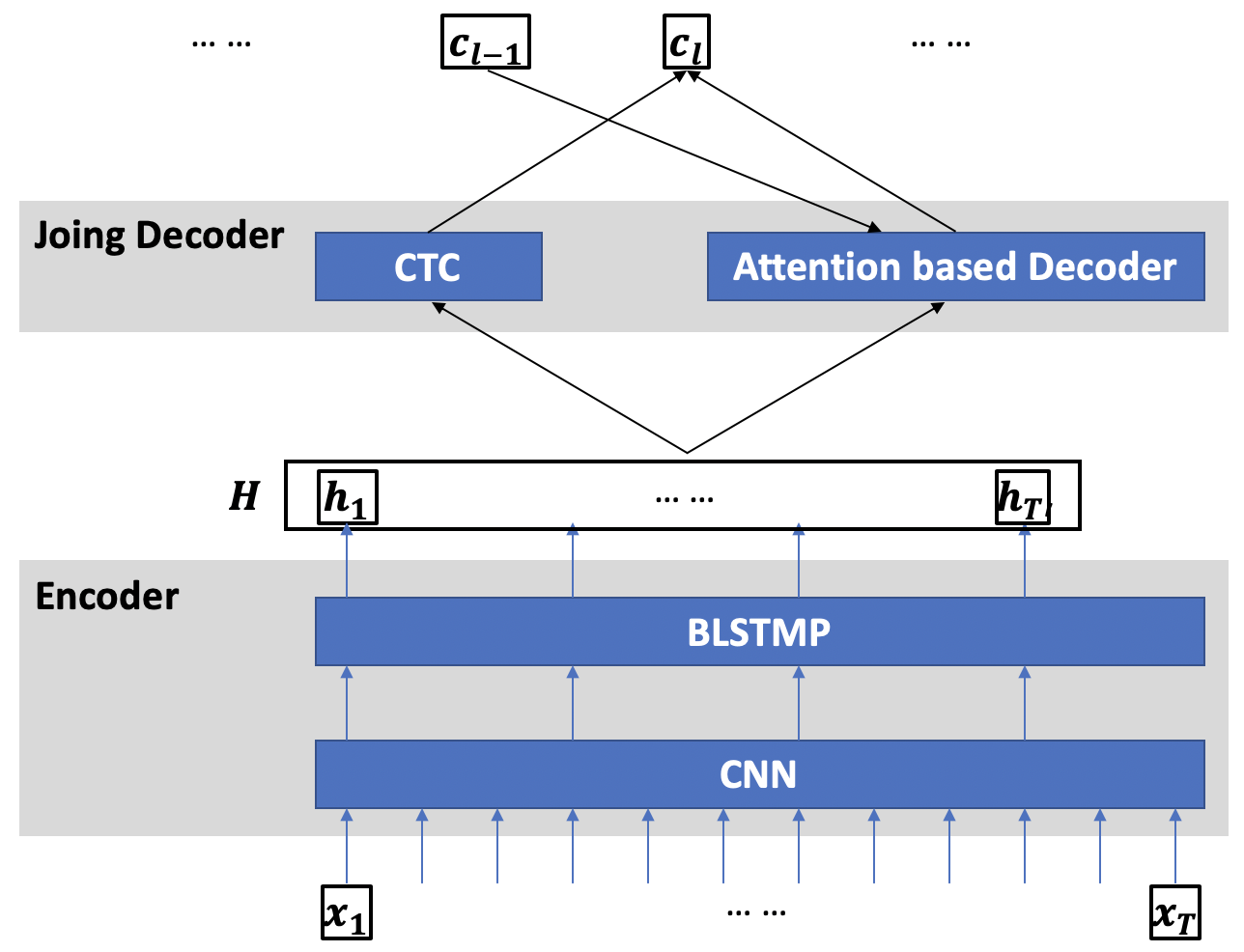}
    \caption{Joint CTC/Attention ASR.}
    \label{fig:las}
\end{figure}

In order to obtain additional benefit on top of the attention-based model, CTC criterion \cite{ctc} was used as an auxiliary task as shown in Fig. \ref{fig:las}.
CTC maximizes the probabilities of the correct L-length label sequence $C= \left \{c_l \in \mathcal{U}\:|\:l=1, ..., L\right \}$ with a set of characters $\mathcal{U}$, conditioned on a given input T-length sequence $X$.
The extra ``blanck" symbol $<$blk$>$ is introduced to map frames and labels to the same length, transforming $C$ to sequence $Z = \left \{z_t \in \mathcal{U} \cup <blk> \:|\:t=1, ..., T\right \}$.
CTC assumes conditional independence between predictions and defines the probability of the label sequence conditioned on the acoustic feature $X$ as follows:
\begin{equation}\label{eq:Pctc}
    P_{ctc}(C|X) = \sum_{Z}^{} \prod_{t}^{} p(z_t | X),
\end{equation}
where $p(z_t | X) = Softmax(Lin(h_t))$ with $h_t = Encoder(X)$.\\
Without independence assumption, an attention-based decoder estimates probability based on the chain rule:
\begin{equation}\label{eq:Patt}
    P_{att}(C|X) = \prod_{l}^{} p(c_l | c_1, ..., c_{l-1}, X),
\end{equation}
where $p(c_l | c_1, ..., c_{l-1}, X) = AttentionDecoder(h, c_{l-1})$ with $h=Encoder(X)$.

For multi-task learning (MTL), logarithms of CTC and attention objectives in Eq. \ref{eq:Pctc} and \ref{eq:Patt} are linearly combined \cite{hybrid_ctc_attention}:
\begin{equation}\label{eq:mtl}
    L_{MTL} = \lambda\:\log{p_{ctc}(C | X)} + (1 - \lambda)\:\log{p_{att}(C | X)},
\end{equation}
with a parameter $\lambda$: $0 \leq \lambda \leq 1$.
We also adopted a joint decoding method which takes the CTC predictions into account during inference \cite{advances_in_joint_ctc_attention}.
In order to combine frame-synchronous CTC probabilities with label-synchronous attention probabilities, we followed the one-pass decoding method described in \cite{advances_in_joint_ctc_attention}, which uses CTC probability during beam search rather than after beam search is done as a rescoring method.
Note that we didn't put an RNN-LM network.

\subsection{Modeling units} \label{ssec:sys_units}
\subsubsection{Korean alphabet} \label{sssec:sys_units_ko}
In the Korean writing system, called Hangul, there are 51 Jamo letters with 30 consonants and 21 vowels just as there are 26 letters each either a consonant or a vowel in English.
\begin{figure}[ht]
    \centering
        \includegraphics[width=5.0cm]{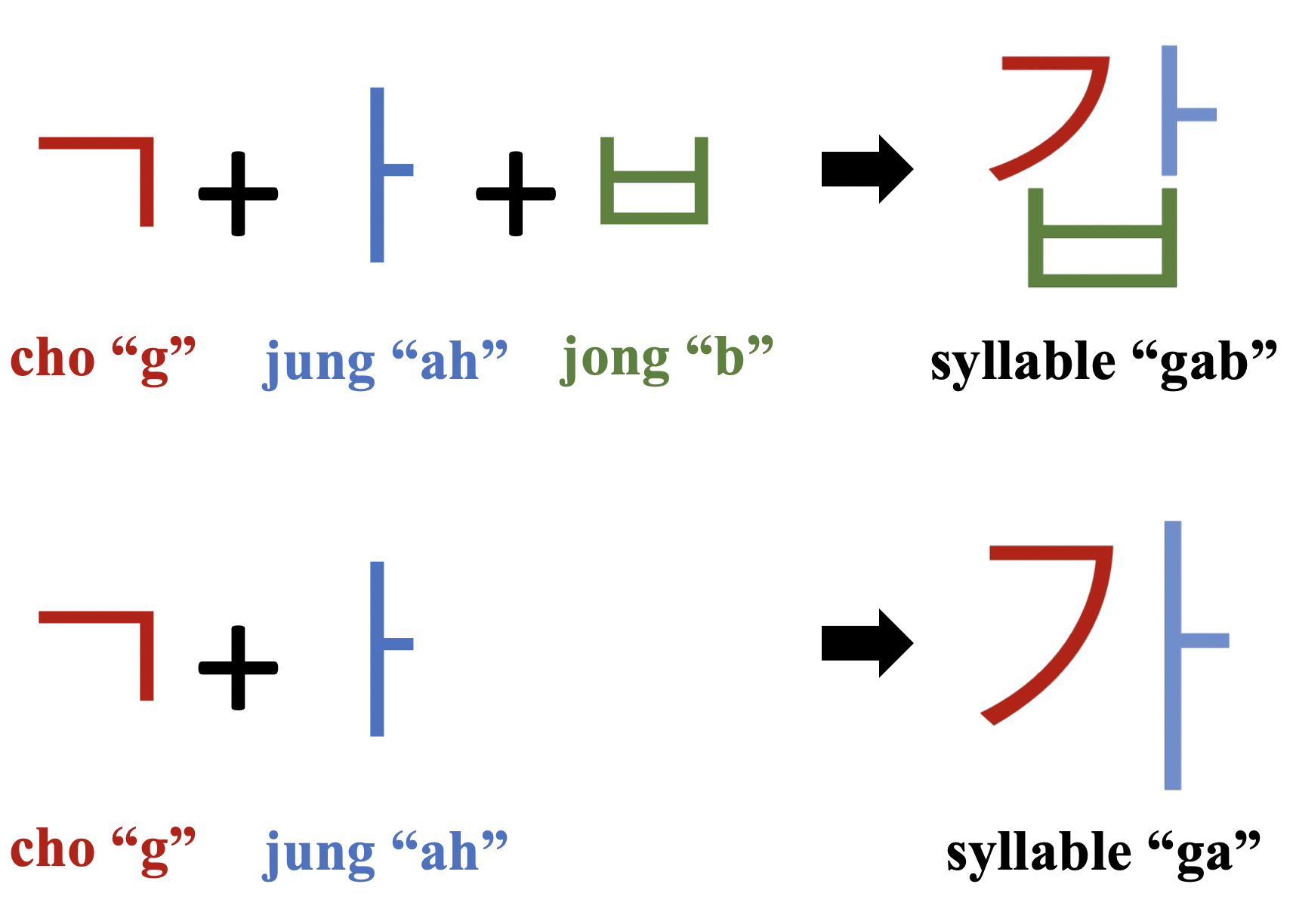}
    \caption{Two or three Jamo letters (left) are grouped forming a syllable block (right), a basic symbol in Korean.}
    \label{fig:syllable}
\end{figure}
Jamo letters are arranged in two dimensions, building a syllabic block which is a basic element of Hangul.
For example, in Fig. \ref{fig:syllable}, first sample shows that three Jamo letters, ㄱ, ㅏ, and ㅂ are combined clock-wisely forming into a syllabic block, 갑.
A single Jamo letter cannot represent elements of the Korean language alone and it must be grouped with others within a block.
A Jamo letter positioned in the first place in a group is called choseong, one in a second position is called jungseong, and third one is called jongseong.
Jongseong is an optional one (bottom of Fig. \ref{fig:syllable}).
A rule for cho, jung, jongseong is as follows: Only 19 consonant letters for choseong, 21 vowels for jungseong, and 28 consonants (including none) for jongseong are available.
This leads to (19 + 21 + 28 = ) 68 units in a Jamo-based system.
Although we can combine Jamo letters to build (19 $\times$ 21 $\times$ 28 = ) 11,172 different Hangul syllables, most of syllables are not used in the real world.
Thus, we only consider top 2350 syllables selected on the basis of frequency.

\subsubsection{Sub-word units} \label{sssec:sys_units_subword}
In order to generate sub-word units, SentencePiece \cite{sentence_piece} was applied on previously described two different units: syllable-based Korean character (right side in Fig. \ref{fig:syllable}) and Jamo characters after breaking up all syllable blocks into cho, jung, jongseong (left in Fig. \ref{fig:syllable}).
Jamo-based sub-word units include Jamo characters, (partial) syllable blocks, and entire word.
For syllable-based sub-word units, they range from a syllable block all the way up to the entire words.
Note that we can choose the number of sub-word units from transcripts in the sentence-piece model.

\begin{table}[ht]
    \centering
    \footnotesize
    \begin{tabular}{@{}c|ccc@{}}
        \toprule
        Units & Examples \\ \toprule
        syllable  &  학, 교, 에, $<$sp$>$, 간, 다 \\ \hline
        jamo & ㅎ,ㅏ,ㄱ,ㄱ,ㅛ,ㅇ,ㅔ,$<$sp$>$,ㄱ,ㅏ,ㄴ,ㄷ,ㅏ \\ \hline
        syll-sw & 학교, 에\underline{\hspace{0.18cm}}, 간, 다 \\ \hline
        jamo-sw  & 학ㄱ,ㅛ, ㅇ, ㅔ\underline{\hspace{0.18cm}},가,ㄴ,다 \\ \hline
        (en) char & I,',m,$<$sp$>$,g,o,i,n,g,$<$sp$>$,t,o,\textless sp\textgreater,s,c,h,o,o,l \\ \hline
        (en) sw & I'm, \underline{\hspace{0.18cm}}go, ing, \underline{\hspace{0.18cm}}to\underline{\hspace{0.18cm}}, s, ch, ool \\ \bottomrule
        (ko) jamo + (en) char & s,c,h,o,o,l,ㅇ,ㅔ,$<$sp$>$,ㄱ,ㅏ,ㄴ,ㄷ,ㅏ\\ \hline
        (ko) syll + (en) char & s,c,h,o,o,l, 에, $<$sp$>$, 간, 다 \\\hline
        (ko) syll-sw + (en) sw & s, ch, ool\underline{\hspace{0.18cm}}, 에\underline{\hspace{0.18cm}}, 간, 다 \\ \hline
        (ko) jamo-sw + (en) sw & s, ch, ool\underline{\hspace{0.18cm}},ㅇ, ㅔ\underline{\hspace{0.18cm}},가,ㄴ,다 \\ \bottomrule
    \end{tabular}
    \caption{Examples of various units for a sample sentence, ``학교에 간다 (I'm going to school)". One of its code-switching version is ``school 에 간다". Token $<$sp$>$ refers to `space'. `sw' means sub-word unit.}
    \label{table:units_ex}
\end{table}
Table \ref{table:units_ex} shows a sample sentence tokenized in different units: syllable, Jamo, syllable-based sub-word, Jamo-based sub-word unit.
Note that in the sentenc-piece model, token $<$sp$>$ is replaced with a token under-bar, which may or may not be joined to another character.
For Korean-English code-switching dataset, we made hybrid units with English characters and sub-word units (bottom of Table \ref{table:units_ex}).

\section{Experiments} \label{sec:exp}
\subsection{Data} \label{ssec:exp_data}
We demonstrate our results on two different ASR corpora, as shown in Table \ref{table:db_stats}.
The first one is called the Zeroth-Korean \cite{zeroth_korean}, which was developed by a Kaldi \cite{kaldi_toolkit}-based Korean ASR open source project called Zeroth project \footnote{https://github.com/goodatlas/zeroth}.
It contains a morpheme-based segmenter called morfessor \cite{morfessor_toolkit} as well as transcribed audio datasets.
We used the morphologically segmented text by using given morfessor model.
Second corpus is our own dataset named as Medical Record (medrec), which consists of large amount of real medical record obtained from Korean hospitals.
It is used for a Korean-English code-switching ASR experiment.
A snippet of medrec is as follows: ``rectal mass 는 이전 보다 volume 이 감소 되고 있음 그러나 여전히 residual tumor mass 는 남아 있음", which is almost half-and-half mixed in Korean and English.
In order to morphologically segment the medrec text data, we used our own rule-based tool rather than the morfessor toolkit.
\begin{table}[ht]
    \centering
    \small
    \begin{tabular}{@{}c|cccc@{}}
        \toprule
        Zeroth & Lang (\%) & Total (h) & Single (s) & \# Spkrs \\ \toprule
        Train & ko 100 & 51.6 & 8 (3 $\sim$ 20) & 105 \\
        Test & ko 100 & 1.19 & 9 (5 $\sim$ 20) & 10 \\
        \bottomrule
        \toprule
        Medrec & Lang (\%) & Total (h) & Single (s) & \# Spkrs \\ \toprule
        Train & ko 40.4 + eng 51.1 & 2530 & 17 (2 $\sim$ 58) & 160 \\
        Test & ko 41.6 + eng 49.2 & 1.16 & 25 (2 $\sim$ 59) & 10\\
        \bottomrule
    \end{tabular}
    \caption{Statistics for Zeroth-Korean (16 kHz, 16 bit) and Medical Record (8 kHz, 8bit) corpora. Column `Single (s)' represents an average duration of a single wave file in seconds.}
    \label{table:db_stats}
\end{table}

We used 80-dimensional log-mel filterbank coefficients with 3-dimensional pitch values extracted following the method described in \cite{pitch}. 
The features are extracted every 10 ms with a 25 ms long hamming window.
And they are normalized using a pre-computed mean and standard deviation value from the training set.

\subsection{Training} \label{ssec:exp_training}
Hybrid CTC/attention architecture is used for all experiments as described in Section \ref{ssec:sys_architecture}.
The number of target units in this model is shown in Table \ref{table:units_labels}.
\begin{table}[ht]
    \centering
    \small
    \begin{tabular}{@{}c|cc@{}}
        \toprule
        Units & \# outputs & Labels \\ \toprule
        syllable & 2371 & syll2350 + \textless sp\textgreater + \textless unk\textgreater + \textless symb\textgreater \\ \hline
        jamo & 88 & jamo68 + \textless sp\textgreater + \textless symb\textgreater \\ \hline
        byte & 256 & 00 $\sim$ ff \\ \bottomrule
    \end{tabular}
    \caption{The number of output classes according to modeling units. \textless symb\textgreater includes spoken symbols such as \#, \%, \&, etc. and numbers such as 0 $\sim$ 9, 10, 100 etc, leading to 19 classes.}
    \label{table:units_labels}
\end{table}
For a unit syllable in the first row, tokens $<$sp$>$, $<$unk$>$, and $<$symb$>$ are added to 2350 syllable characters.
Only $<$sp$>$ and $<$symb$>$ are added to 68 Jamo characters since an unknown token does not exist in a jobo-based system.
For Korean-English code-switching ASR task, we added 26 English characters and one apostrophe symbol, leading to 27 additional classes.
For a byte unit \cite{bytes_are_all_you_need}, no extra labels were added other than 256 units.
For sub-word units from the sentence-piece model, we generated 3k and 6k target sets on a syllable-based text, 2k and 3k sets on a Jamo-based text.
Note that two common tokens, $<$blk$>$ for CTC and $<$sos/eos$>$ (start/end of sentence) for attention-based decoder, are added for all modeling units including sub-word sets.
The parameters of the model are about 39 M to 48 M according to the number of target labels.

During training, we kept a value of MTL weight parameter $\lambda$ in Eq. \ref{eq:mtl} as 0.2.
Training and evaluation were done using Espnet toolkit \cite{espnet_toolkit} and Chainer CTC was used.
During training, unigram label smoothing was employed as described in \cite{state_of_the_art_seq2seq}.
The Adadelta \cite{adadelta} with gradient clipping was used for the optimization.
For a beam search during inference, the beam width was set to 30.
The implemented decoding strategy is described in Section \ref{ssec:exp_training}.

\section{Results and discussion}
\label{sec:results}
We compared five different modeling units, which are syllable, Jamo, syllable-based sub-word, Jamo-based sub-word, and a byte.
Table \ref{table:results_zeroth} and \ref{table:results_medrec} summarize the ASR results in terms of character / word / sentence error rate (CER / WER / SER) on Zeroth-Korean (only Korean) and Medical Record (Korean-English code-switching), respectively.
\begin{table}[ht]
    \centering
    \small
    \begin{tabular}{@{}c|ccc@{}}
        \toprule
        Units & CER (\%) & WER (\%) & SER (\%) \\ \toprule
        syllable & 1.8 & 2.6 & 3.3  \\
        jamo & 6.3 & 19.9 & 83.6 \\
        syll-subword (3k) & 3.0 & 3.2 & 5.0 \\
        syll-subword (6k) & 2.3 & \textbf{2.5} & 3.3 \\
        jamo-subword (2k)  & 75.3 & 4.1 & 4.8 \\
        jamo-subword (3k)  & 4.2 & 4.3 & 4.4 \\
        byte & 2.5 & 4.4 & 13.1 \\
        \bottomrule
    \end{tabular}
    \caption{CER, WER, SER (\%) on test set of Zeroth-Korean with different modeling units.}
    \label{table:results_zeroth}
\end{table}

According to Table \ref{table:results_zeroth}, the syllable-based sub-word unit achieved the best performance while Jamo unit resulted in the highest WER value for the Korean ASR system.
As Jamo units are grouped by a sentence-piece model, the performance gets better as you can see in the second, fifth and sixth row.
The results imply that the modeling units with a longer scale such as syllables and sub-words outperform ones with a shorter scale such as Jamo.
\begin{figure}[ht]
    \centering
        \includegraphics[width=5.8cm]{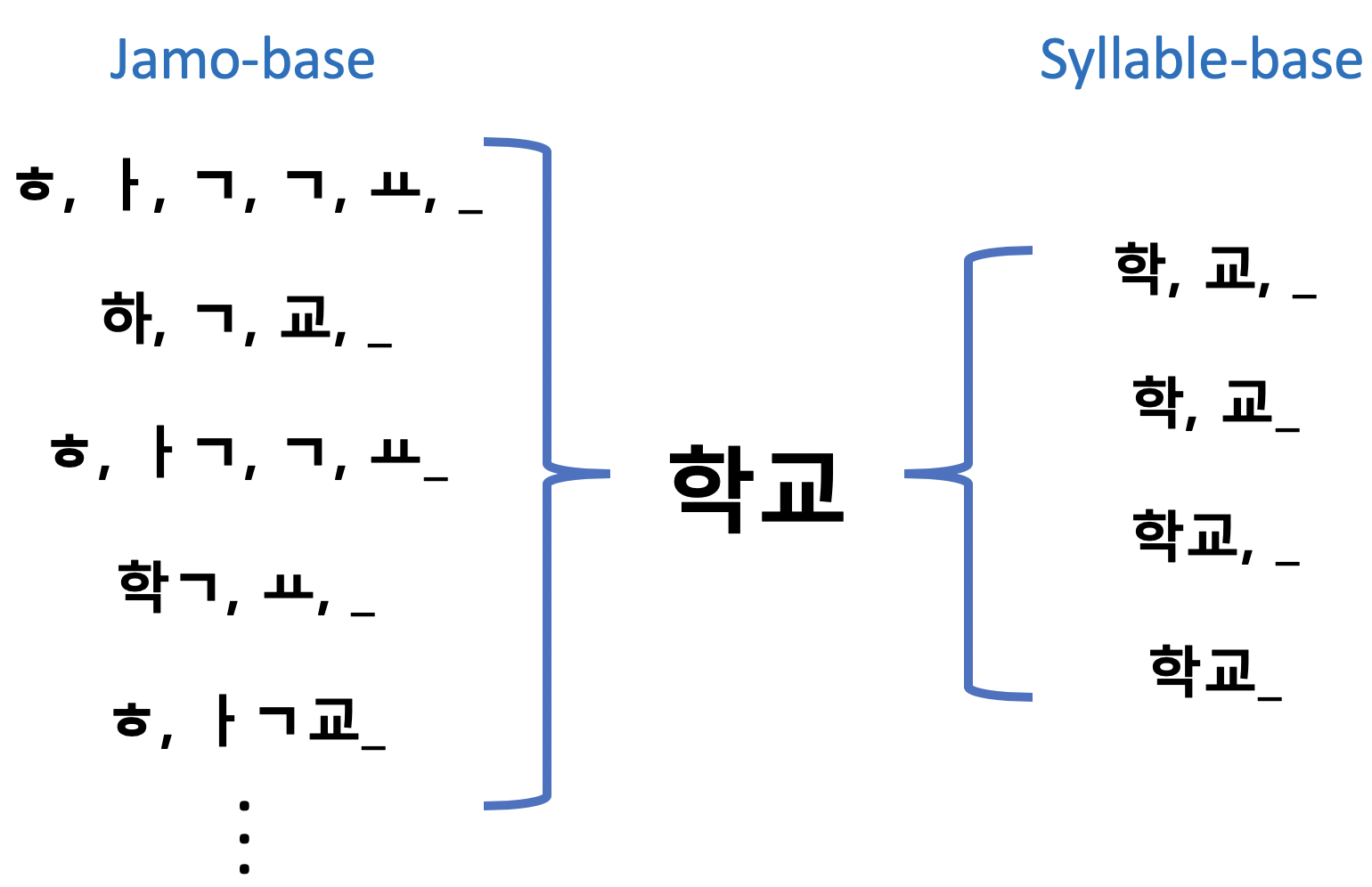}
    \caption{A sample word `학교', meaning `school' in English. Word `학교' can be encoded into various Jamo-based sub-word units as shown in the left part. For syllable-based sub-word units on the right side, only 4 cases are possible.}
    \label{fig:haggyo}
\end{figure}
If you see the fifth row in Table \ref{table:results_zeroth}, Jamo-based sub-word units result in a high CER value, 75.3 \% while achieving low WER value of 4.1 \%.
The reason is described in Fig. \ref{fig:haggyo}.
For a Jamo-based unit, there are a number of possible combinations (left side) collapsing into a single word unlike a syllable-based unit system which has only several combinations (right side).
Such a different representation for the same word in a Jamo-based system leads to a high CER value.
This kind of problem is reduced as more grouping of Jamo letters is done (sixth row).

Jamo has a similar scale with English characters while the syllable unit has a longer scale.
In a code-switching task (Table \ref{table:results_medrec}), however, a syllable-based Korean character with an English character was a better choice than the combination of Jamo and English character (first and second row).
The best modeling units in this task were found to be the combination of the syllable-based sub-word unit and the English sub-word unit (fourth row), which is consistent with results of the Korean speech recognition in Table \ref{table:results_zeroth}.
Note that byte unit \cite{bytes_are_all_you_need}, universal one across any languages, reached lower WER value than Jamo unit for both tasks.
\begin{table}[ht]
    \centering
    \small
    \begin{tabular}{@{}c|ccc@{}}
        \toprule
        Units & CER (\%) & WER (\%) & SER (\%) \\ \toprule
        (ko) syllable + (en) char & 4.6 & 8.1 & 66.1  \\
        (ko) jamo + (en) char & 6.7 & 16.8 & 92.3 \\
        syll-subword (3k) & 9.7 & 8.2 & 65.5 \\
        syll-subword (6k) & 7.7 & \textbf{6.9} & 64.2 \\
        byte & 6.4 & 10.4 & 77.0 \\
        \bottomrule
    \end{tabular}
    \caption{CER, WER, SER (\%) on test set of Medical Record with different modeling units.}
    \label{table:results_medrec}
\end{table}

\section{Conclusions and future work}
\label{sec:conclusion}
In this work, we presented a comparative study of different modeling units in both Korean and Korean-English code-switching ASR applications.
Investigated units include syllable-based Korean character, Jamo, and syllable/Jamo-based sub-word units.
Our experiments using a CTC/attention-based sequence-to-sequence model showed that the sub-word unit based on Korean syllables performed the best, which is a consistent result with a code-switching task.
In the future, we would like to compare the results with Transformer \cite{speech_transformer}, another competitive sequence-to-sequence model architecture.

\vfill\pagebreak

\bibliographystyle{IEEEbib}
\bibliography{papers}

\end{document}